# Thermal-driven Flow inside Graphene Channels for Water Desalination

Bo Chen, Haifeng Jiang*, Huidong Liu, Kang Liu, Xiang Liu & Xuejiao Hu*

Key Laboratory of Hydraulic Machinery Transients (Wuhan University), Ministry of Education, School of Power and Mechanical Engineering, Wuhan University, Wuhan, Hubei 430072, China

*Corresponding authors. E-mail addresses: hfjiang@whu.edu.cn (H. Jiang), xjhu@whu.edu.cn (X. Hu).

**ABSTRACT**

A novel concept of membrane process in thermal-driven system is proposed for water desalination. By means of molecular dynamics simulations, we show fast water transport through graphene galleries at a temperature gradient. Water molecules are driven to migrate through nanometer-wide graphene channels from cold reservoir to hot reservoir by the effect of thermal creep flow. Reducing the interlayer spacing to 6.5 Å, an abrupt escalation occurs in water permeation between angstrom-distance graphene slabs. The change from disordered bulklike water to quasi-square structure have been found under this extremely confined condition. This leads to a transition to subcontinuum transport. Water molecules perform collective diffusion behaviors inside graphene channels. The special transport processes with structure change convert thermal energy into motion without dissipation, resulting in unexpected high water permeability. The thermal-driven system reaches maximum flowrate at temperature variance of 80 K, corresponding to the quantity at pressure difference up to $10^5$ bar in commercial reverse osmosis processes and 230 bar in pressure-driven slip flow. Our results also reveal the movement of saline ions influenced by thermophoretic effect, which complement the geometry limitation at greater layer spacing, enhancing the blockage of ions. This finding aims to provide an innovational idea of developing a high-efficiency desalination technology able to utilize various forms of energy.

The shortage of water resources have been highly severe in recent years due to ever-increasing global population, rapid industrialization and expanding urbanization.[1-4] Desalination of saline water has now been accepted as a promise way to gain fresh water.[5-8] Well-established desalination technologies can be classified as membrane processes (without phase change) and thermal processes (with phase change). Membrane processes utilize a physical barrier (membrane) to separate the dissolved ions from the feed water, with low consumption of energy and high efficiency of water product, but membrane-based technologies such as reverse osmosis (RO) processes usually demand high-grade energy (electric energy).[7-11] Thermal processes vaporize freshwater from the saline water, so various forms of energy, such as fossil fuel sources, solar energy, or waste heat, can be directly used in thermal desalination technologies.[12-16] However, the great latent heat of vaporization makes the thermal processes energy-extensive consumption.[1, 6] Those limitations of membrane processes and thermal processes have made the development of desalination technologies stalled.

Nowadays, the blooming of advanced materials in nanoscale applied in desalination brings many chances to get out of the dilemma.[17-24] Water interlaminated between two-dimensional (2D) materials, such as graphene oxide (GO) membranes, is proved to show unique flow behaviors and unconventional mechanisms. Unimpeded permeation of water through GO membranes were experimentally investigated in many researchers, and it is found that the permeation rates of water are several orders of magnitude greater than the classic viscous-flow prediction.[25-28] This unexpected fast flow was attributed to the slippery nature of water flow in confined channels with great Knudsen number, and the pristine graphene regions in GO sheets offer the super-lubrication boundary to enhance the slip effect.[29-36] Recent experimental and theoretical studies have demonstrated the existence of ordered water structures within nanoconfined space.[37-39] This solid-like ordered water performs collective diffusion, which causes fast water transport through GO membranes at high humidity.[40] The collective behaviors of water molecules with ordered structures shows different characteristics at various temperature, which can cause a fountain flow of water across nanotubes with temperature gradient.[41] Though the nano-scaled materials up to date have improved the membrane performance to some extent, it still depends on the pressure-driven system for operating. Is there some way to combine the high-efficiency of membrane processes with the wide-use of energy of thermal processes?

In the present work, we propose a new concept of membrane process for water desalination in thermal-driven system, through performing molecular dynamics (MD) simulations. It may help to develop an innovative method for high-efficiency desalination, which is able to utilize various forms of energy. The molecular structure is shown in Figure 1. Two saline water reservoirs are separated by a graphene channel, and a heat current is applied to the whole system, maintaining the two reservoirs at different temperature. We have found that water molecules are driven to transport from the cold side to hot side by the effect of thermal creep flow, while saline ions are driven to move at reverse direction owing to the thermophoretic force. These two effects help water separate from saline water out of pressure-based system. What's more, a subcontinuum transport has been revealed in extremely confined water molecules, exhibiting collective diffusion behaviors and convert thermal energy into directed motion with high efficiency. It shows extraordinary ability of water permeation, corresponding to that of commercial RO membrane at pressure difference of $10^5$ bar and 230 bar in pressure-driven slip flow.

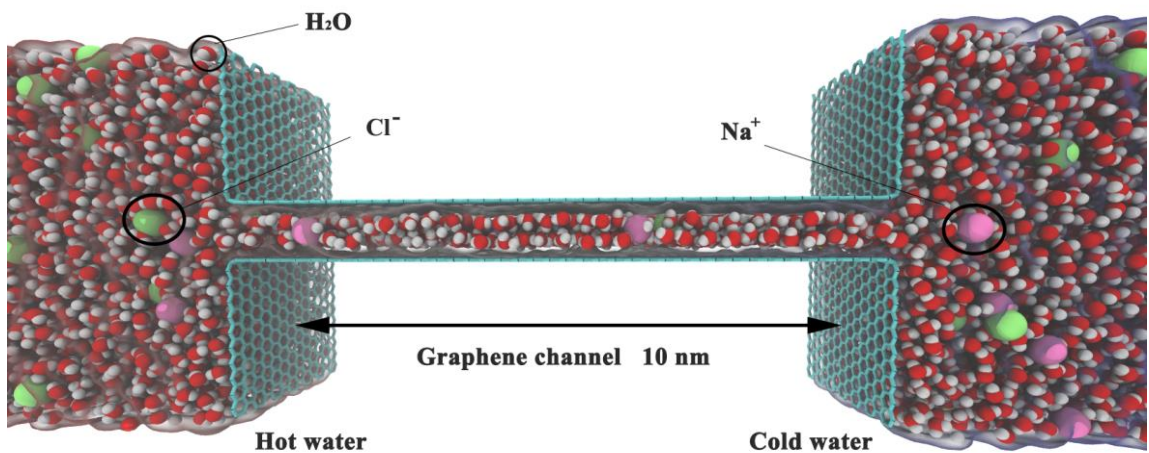

**Figure 1. The diagram of thermal-driven flow through graphene channel. Cyan, red, white, lime and mauve spheres represent carbon atoms, oxygen atoms, hydrogen atoms, sodium ions, and chloride**

**ions respectively.**

**Thermal-driven Creep Flow.** The temperature field plays a significant role in inducing flow, especially for great Knudsen numbers. According to the rarified gas theory, a net flow can be driven along a temperature gradient on a diffusely reflecting wall, which is known as thermal creep flow.[42-46] For the flow confined between two flat plates with a temperature gradient $\partial T/\partial x$, the induced velocity $u$ can be estimated as,[47-48]

$$u = \frac{3+j}{4}\left(\frac{\eta}{\rho T}\frac{\partial T}{\partial x}\right)_s. \tag{1}$$

Here, $j$, $\eta$, $\rho$, and $T_s$ is the internal degree of freedom, viscosity, density, and temperature of fluid molecules, respectively. We consider the averaged velocity $u_m$ along the whole channel to assess the comprehensive thermal creep flow, which is defined as:

$$u_m = \frac{1}{L}\int_0^L \frac{3+j}{4}\left(\frac{\eta}{\rho(T_0+\frac{\partial T}{\partial x}x)}\frac{\partial T}{\partial x}\right)_s dx = \frac{3+j}{4}\cdot\frac{\eta}{\rho L}\cdot\ln(1+\frac{L}{T_0}\frac{\partial T}{\partial x}). \tag{2}$$

Here, $L$ is the length of the channel and $T_0$ is the temperature of the channel entrance. The explanations of thermal creep flow in detail have been expounded in Supplementary Information Section 1 (Figure S1). To elucidate the thermal-driven creep flow inner graphene channel, a passage with $d$-spacing of 12 Å is investigated through the thermostatic system, with cold reservoir at 293 K and hot reservoir at 323 K. Figure 2(a) portrays the temperature distribution of water on $x$-$z$ plane and along $z$ direction. The temperature is linearly increased from cold reservoir to hot reservoir through the graphene channel, promising a stable temperature gradient during the simulation. It is noted that the geometrical spacing between two graphene sheets is greater than the actual flowing width of the channel, because of the repulsive interaction between water molecules and carbon atoms in graphene. The actual width of channel is measured as 8 Å according to the temperature contour figure. The flowrate of water through the graphene channel is considered with the effective flow area, which can be calculated via the product between actual width of channel and the slabs length in *y* direction. The whole simulation box is divided as three parts by the graphene channel, and the number variation of water molecules in each portion has been tracked in Figure 2(b). It is clearly that the water molecules move from cold reservoir to hot reservoir through the graphene channel. The flow direction is identical to the way of positive temperature gradient, consistent with the prediction of thermal creep flow in equation (1). The flow profiles show that the flow rate of water is stable in time, corresponding to 1916 L/cm$^2$/day. Surprisingly, the water flowrate driven from a temperature difference of 30 K is equivalent to that of the existing commercial RO membranes from a pressure difference of almost 10$^4$ bar.

To give a further insight into the flow characteristics, we investigate the thermal creep flowrate at different temperature gradient by setting series of temperature difference from 40 K to 80 K (Figure S2 to Figure S9). The flowrates as a function of temperature difference are concluded in Figure 3. For graphene channels with *d*-spacing beyond 9 Å, the flowrate rises with the increase of temperature difference, but the growth rate of flowrate shows decay at large temperature difference. It accords with the tendency of napierian logarithm as portrayed in the solid curves (Figure 3), consistent with the prediction in equation (2). The flowrate also varies a lot among different graphene channels. That with smaller *d*-spacing exhibits superior capability for thermal-driven flow. It is noted that the graphene channel with *d*-spacing of 6.5 Å shows distinctive flow characteristics compared to others. The flowrate through the narrowest graphene channel greatly exceeds that beyond 6.5 Å and it increases linearly without decays even at larger temperature difference. All above prove that the thermal creep flow is strongly related to the spacing inner graphene channels, and it is important to figure out the size effects on water flowrate.

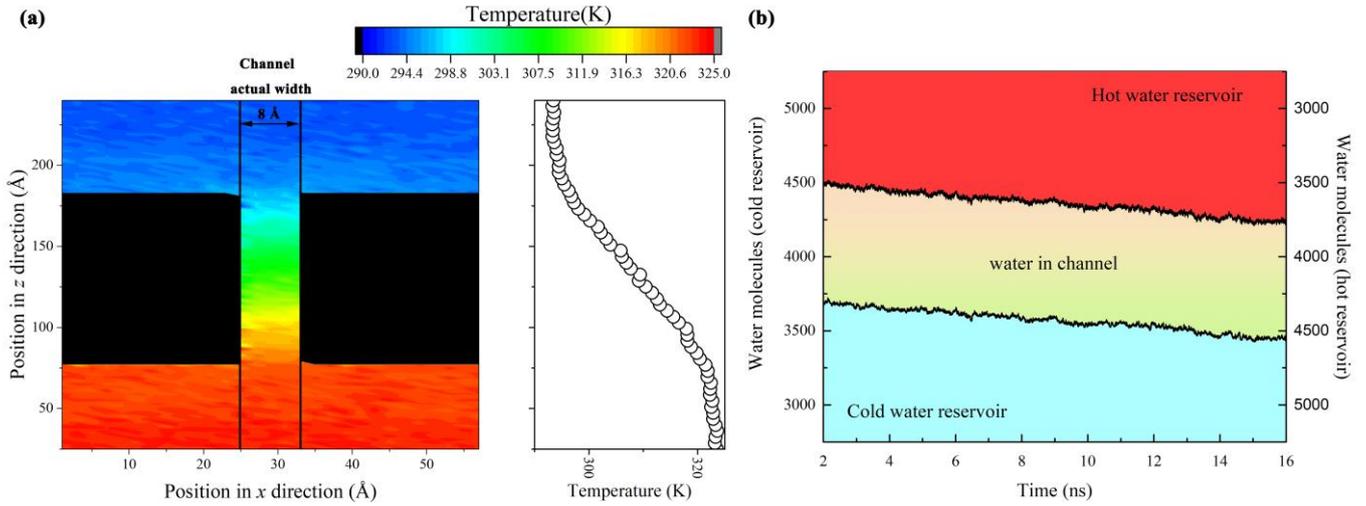

**Figure 2. Thermal-driven creep flow in graphene channel with *d*-spacing of 12 Å. (a) Temperature distribution of water on *x-z* plane (left panel) and along *z* direction (right panel). (b) Temporal evolution of water molecules.**

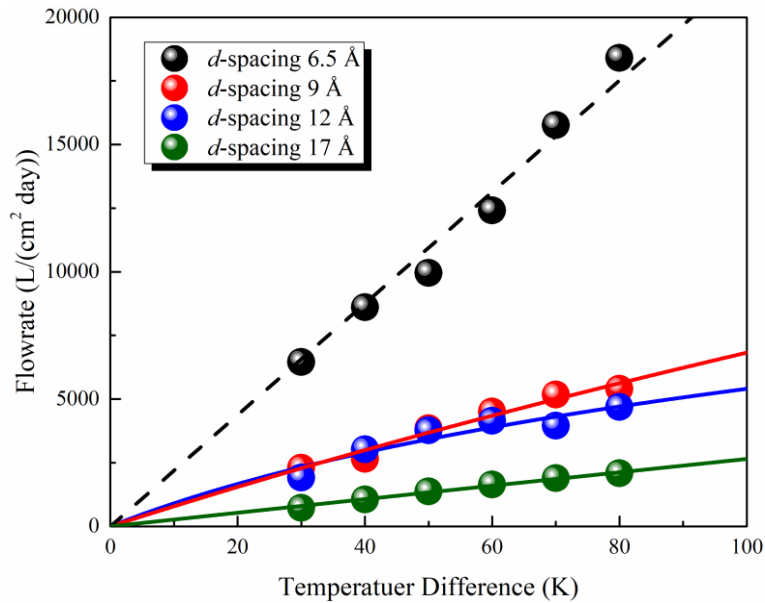

**Figure 3. Dependence of water flowrate through graphene channel with variable *d*-spacing of temperature difference. Black, red, blue, and green spheres represent the value of *d*-spacing corresponding to 6.5 Å, 9 Å, 12 Å, and 17 Å. Lines are guides to the trend of flowrate evolution. Dashed straight line denotes the linear trend, and solid curve lines denote the trend of napierian logarithm with the form of $y = A\ln(1+Bx)$.**

**Size Effects: Transition to Subcontinuum Transport.** Variable kinds of graphene channels with $d$-spacing from 6.5 Å to 22 Å at $\Delta T$ = 30 K have been performed to investigate the size effects on water thermal creep flow (Figure S10-S12). In graphene channels with $d$-spacing larger than 9 Å, where a continuum description of mainstream flow is still valid, we found that the flowrate monotonically decreases with increasing layer spacing (Figure 4). This trend is like the decreasing flow enhancement of pressure-driven flow in carbon nanotubes (CNTs) with growing diameter, due to the impair size effect for large $d$-spacing. It is interesting that a growth spurt occurs when the layer spacing reduces from 9 Å to 6.5 Å, which suggests a transition to subcontinuum transport (Figure 4). By reducing interlayer spacing, the water structure changes from bulklike to highly ordered as shown in the snapshots inserted in Figure 4. To clearly exhibit the difference of water structures and flow characteristics at variable $d$-spacing, we implement simulations of a sandwiched structure with water molecules intercalated between two graphene sheets (Figure S13). We provide the pair correlation functions atoms in water molecules inner graphene gallery with different spacing (Figure S14). The first peak of $g$(O-O) and $g$(O-H) curves vanish with the increase of $d$-spacing, which gradually changes toward the shape of bulk water. This transformation indicates that the scale effect on water inner graphene gallery is progressively weakened with $d$-spacing increasing from 9 Å to 22 Å. To quantify the variation in nanoconfined flow with layer spacing, we perform the calculation of total energy interaction between water molecules and graphene laminations, which is the derivation of thermal creep flow. The averaged energy applied to every water molecule reduces with expanding layer space as shown in Figure S15. It arises from the fact that the intercalated water molecules increases fast while the contact area is limited for channel with a certain length.

When the layer distance dwindle to the extent which only can hold mono-layer water molecules, the structure of water shows significant difference compared to others. The first peak of g(O-O) curves in $d$-spacing of 6.5 Å is much higher than that beyond, and another peak occurs after the first peak. As portrayed in the snapshots of Figure 5(c), it will form monolayer structured water with quasi-square lattice in the extreme confined space (i.e. inner graphene gallery with the layer distance of 6.5 Å). The water with ice-like structure performs collective behaviors to transport through graphene channel in high efficiency. We investigate the in-plane displacement within 30 ps of water molecules under two graphene slabs (Figure S13). In contrast to the diffusion of bulk water, the displacement is decomposed in $x$ and $y$ directions, and the distribution of relative frequency in each quantity of diffusion distance have been calculated for 6.5-angstrom-wide and 22-angstrom-wide channel at 300 K (Figure 5b). It is observed that the diffusion along the $x$ and $y$ directions are anisotropic inside channel with the $d$-spacing of 6.5 Å, while it turns to isotropic inner channel with greater space. The collective diffusion coefficients of intercalated water have also been investigated from the trajectories of center of mass of molecules (Figure 5a). Below the temperature of 310 K, the coefficient $D$ in $x$ direction exceeds that in $y$ direction, which proves that the collective diffusion performs auxo-action on water transport through graphene channel. However, this regular structure no longer exists as increase of $T$, which can be presented in the peak decay of correlation curves (Figure S14). What's more, the collective diffusion behaviors do not predominate in $x$ direction through the surpassing diffusion coefficient in $y$ direction beyond 310 K (Figure 5a). The distribution of diffusion distance presents the trend of anisotropy−isotropy transition in greater temperature (Figure 5b). These results indicate that the ordered quasi-square structure is sensitive to temperature, and it gets collection at low temperature while collapse at high temperature. In the thermal-driven system with $\Delta T$ = 30 K, water molecules are in different temperature at the entrance (293.15 K) and the outlet (323.15 K) of graphene channel. It makes a transition of water structures from ice-like ordered to free along the direction toward positive temperature gradient. Water molecules entering the graphene gallery form an ordered structure and release heat to cold reservoir. Then ice-like water performs a collective transport through the graphene channel to the hot water reservoir. When arriving at the outlet, ordered water molecules gradually transform to free and absorb heat from hot reservoir. The diagram of molecular mechanism is portrayed in Figure 5(d). The collective transport of water molecules inner gallery with low energy dissipation and the directly transformation from heat energy into molecular motion, provide more directional and higher efficient transport compared to the diffusion in bulklike water. Thus, the graphene channel with *d*-spacing of 6.5 Å can achieve such high flowrate.

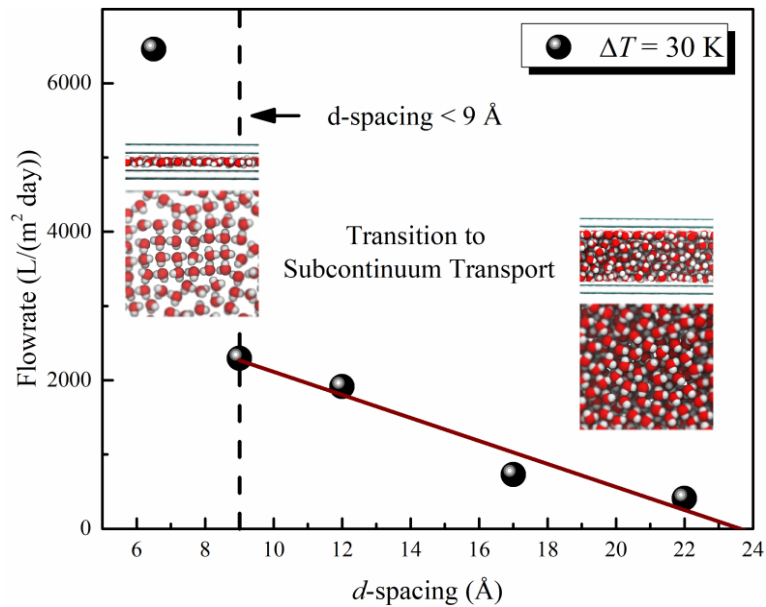

**Figure 4. Dependence of water flowrate through graphene channel of variable *d*-spacing. The inserted figures are the snapshots of water molecules inside graphene channels. Left one portrays the highly ordered water structure confined between graphene channel with *d*-spacing of 6.5 Å, and right one portrays the bulk water structure between graphene channel with *d*-spacing of 12 Å.**

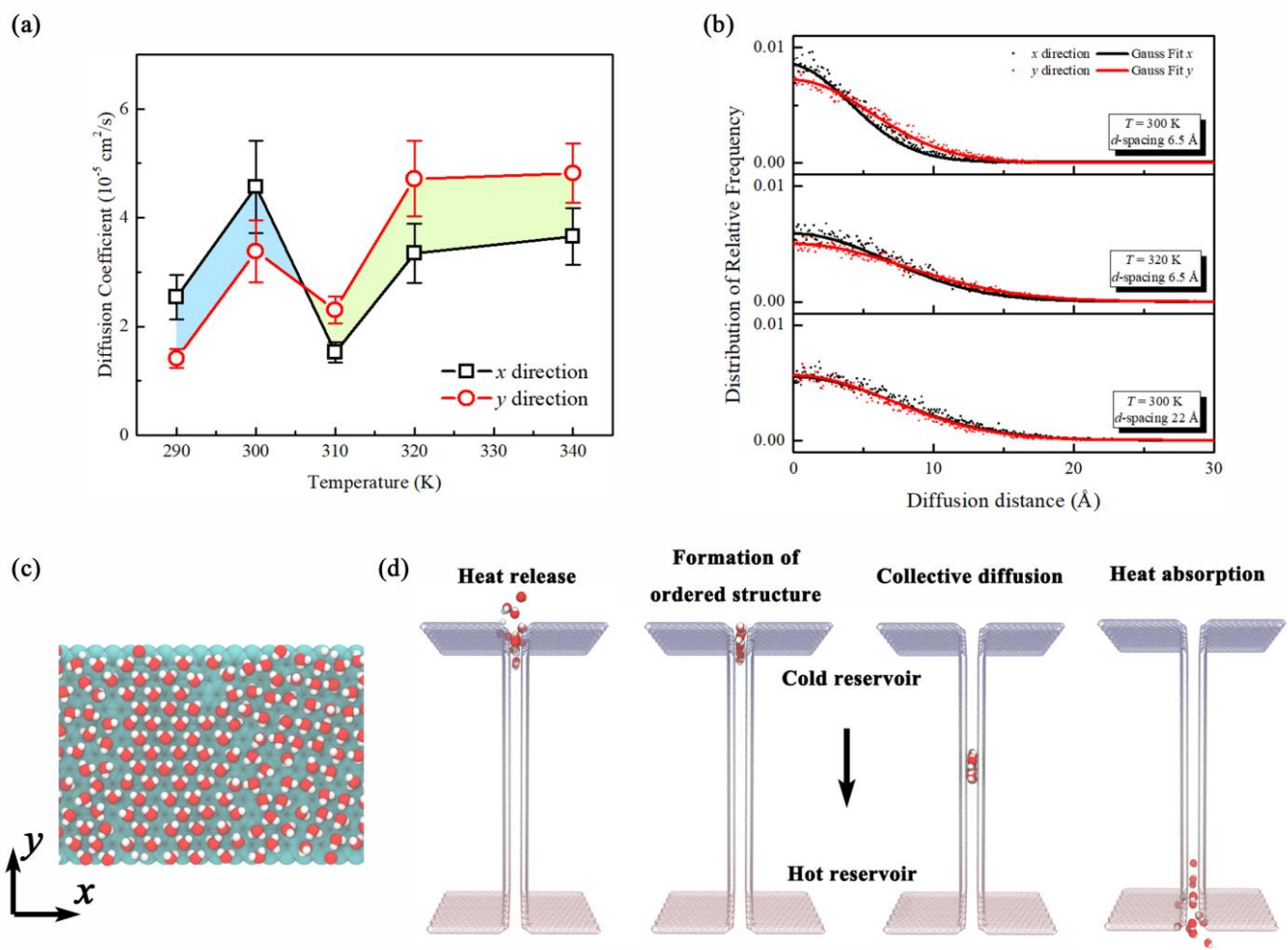

**Figure 5.** (a) Dependence of collective diffusion coefficients of water flow through 6.5-angstrom-wide graphene channel to temperature. (b) Distribution of molecular in-plane displacements within 30 ps along the *x* and *y* directions. (c) Snapshots of ordered structure of intercalated water confined between two graphene sheets. (d) Molecular mechanism for subcontinuum transport of water.

**Thermophoretic effect on ions.** In pressure-driven system, ions filtration is usually dependent on the geometry designation of pores or channels on membranes, which should be smaller than the size of hydrated ions. The hydration radii of some monovalent salt ions (like $Na^+$ and $K^+$) are always at sub-nanometer, so most 2D materials have to downsize the *d*-spacing between laminations at the value not exceeding 0.8 nm to obtain an acceptable salt rejection. Saline water can be considered as the fluid mixture of several large particles (hydrated ions) around many tiny particles (water molecules). When the system is exerted in a temperature gradient, the large particles receive imbalanced collisions from tiny particles in hot sides and cold sides with different momentum. The comprehensive result is a thermophoretic force applied to large particles (saline ions), which get a drift along the negative temperature gradient. It is reverse to the direction of water flow. Thus, there is an additional thermophoretic effect to help impeded ions besides geometry filtration in thermal-driven system. To explicitly indicate the dynamic motion of ions affected only by thermophoretic force, we construct the saline water reservoir at the same ionic concentration. For graphene channel with layer distance of 6.5 Å, smaller than size of hydrated ions, it presents the ability of blocking saline ions absolutely (Figure S16), owing to the size filtration of geometry. As for larger $d$-spacing of 12 Å, the ionic concentrations of $Na^+$ and $Cl^-$ declines in hot reservoir (permeation side), and in contrast, those increase in cold reservoir (feeding side) during the whole process of thermal-driven flow. It proves that water in hot reservoir can be purified even on the weak effect of geometry. To explore the motions of ions inner graphene channel, we track the trajectories of some representative ions for 10 ns. It can be seen that ions hesitate in the gallery and finally migrate from hot reservoir to cold reservoir. It is because that two external forces, viscous stress by relative shift of water flow and thermophoretic force by temperature gradient, are both exerted to saline ions. At temperature difference corresponding to 30 K, the major factor is the thermophoretic effect, resulting in the ionic motion from hot side to cold side.

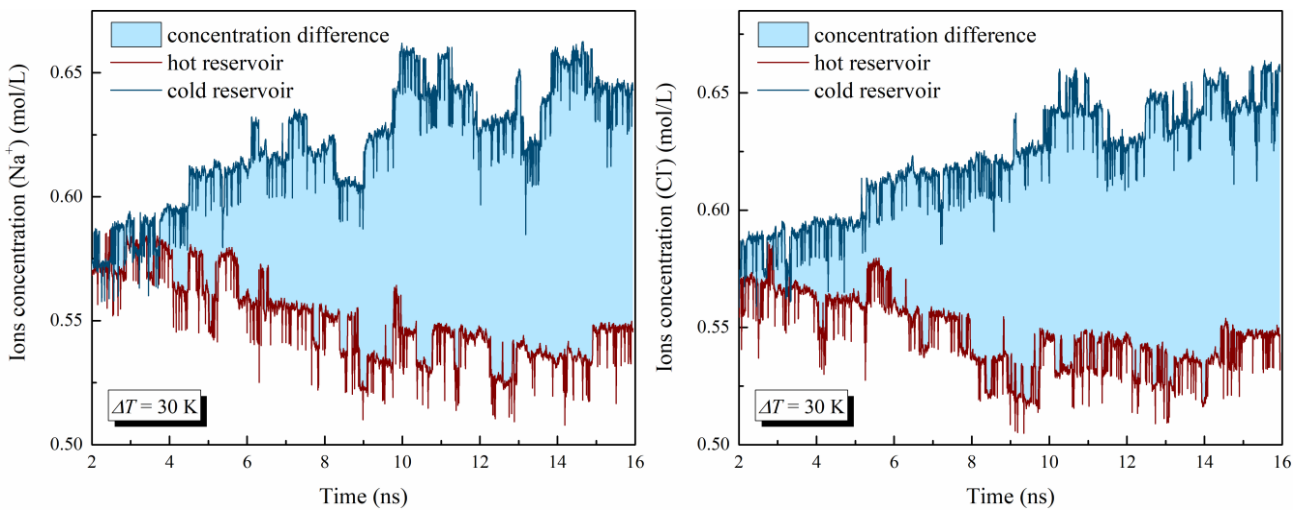

**Figure 6.** Temporal evolution of ions concentration in the case of graphene channel with *d*-spacing of 12 Å. Left panel portrays the ions concentration of Na$^+$, and right panel portrays that of Cl$^-$.

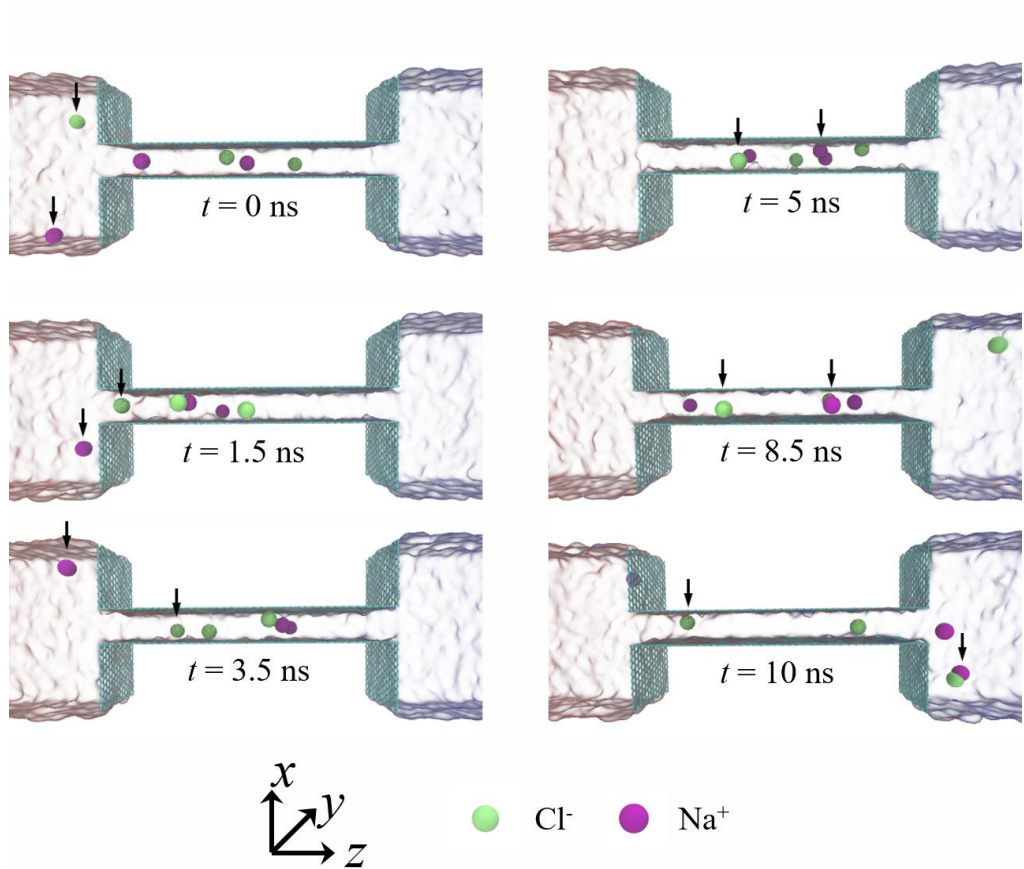

**Figure 7.** Snapshots of representative ions transport through graphene channel with *d*-spacing of 12 Å. Green particles denote Na$^+$, and purple particles denote Cl$^-$. Left side represents hot reservoir colored by red, and right side represents cold reservoir colored by blue. The black arrows are guide to eye for the same ions.

**Comparisons with pressure-driven flow.** Finally, we make a comparison of water flowrate between thermal-driven system and pressure-driven system. According to MD results, graphene channels under temperature gradient exhibit remarkable permeability of water molecules, with a flowrate of 732 ~ 18402 L/cm$^2$/day. This quantity corresponds to the infiltration capacity of existing commercial membranes at the pressure difference of $10^4$ to $10^5$ bar, as shown in the blue arrows of Figure 8. To complement the pressure-driven flow, we perform the simulations of water flow through the same molecular structures with $d$-spacing of 6.5 Å at 192 bar. Fast slip flow has been found for water inside the interlayer gallery between graphene slabs, which has also been investigated in previous works. We linearly extrapolate the flowrate at different pressure differences as plotted in the purple straight line (Figure 8), through Darcy's law, $\bar{v} = \gamma\left(\frac{\Delta P}{L}\right)$. Here, $\Delta P$, $L$ and $\gamma$ is the pressure difference across the channel, the channel length, hydraulic conductivity, respectively. The flowrate of graphene galleries with $d$-spacing of 17 Å can attain the extent of slip flow at $\Delta P = 10 \sim 30$ bar, and with $d$-spacing of 9 Å to 12 Å it reaches the extent of slip flow at $\Delta P = 20 \sim 70$ bar. The maximum flowrate is obtained through the graphene channel with layer distance of 6.5 Å. A surprising flowrate at pressure difference up to 230 bar in slip flow is equal to that in thermal-driven system at a temperature difference of 80 K. Overall, our results indicate that the thermal-driven creep flow through graphene channels could act as a high-permeability system for water desalination.

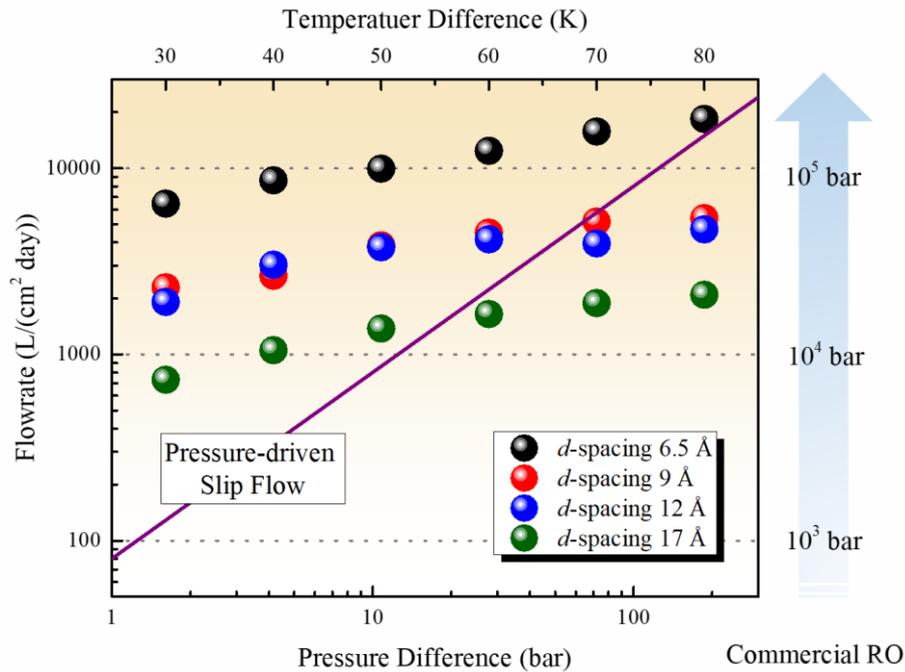

**Figure 8. Comparisons of water flowrate with pressure-driven flow. The scatters are MD results of**

**thermal-driven flow with top *x*-axis in temperature difference. The straight line is linearly deduced from simulations of pressure-driven flow with slippage boundary with bottom *x*-axis in pressure difference. The grey dotted lines represent the required pressure of commercial RO to achieve such flowrate.**

In summary, we propose a novel membrane process for water desalination in thermal-driven system, through nanoscale graphene channels. Our MD simulations results show that water molecules are driven to migrate through nanometer-wide graphene galleries from cold reservoir to hot reservoir by the effect of thermal creep flow. It derives to the imbalanced momentum exchange between graphene slabs and water molecules from two reservoirs. The flow velocity is strongly related to the temperature gradient and get an increasing trend as the form of napierian logarithm with greater temperature difference, corresponding to the prediction of kinetic theory. The size effect is another significant factor on water flow. The flowrate monotonically decreases with increasing layer spacing beyond 9 Å, showing impair size effect for greater $d$-spacing. The abrupt escalation in water permeation between the 9 and 6.5 angstrom-distance graphene slabs suggests a transition to subcontinuum transport, which is consistent with the change from disordered water to quasi-square structure with decreasing $d$-spacing. The collective diffusion of water molecules inside graphene channels and special transport processes with structure change lead to the unexpected high water permeability through graphene galleries. Movement of saline ions are also influenced by the system with temperature gradient, driven to move at a reverse direction of water flow owing to the thermophoretic force. On one hand, the graphene channels could reject ions by geometry limitation ($d$-spacing of 6.5 Å), and on the other hand, it could enhance the blockage of ions by an additional thermophoretic effects even at greater layer spacing. This thermal-driven system shows extraordinary performance compared to existing commercial RO membrane and pressure-driven slip flow. It get a maximum flowrate through 6.5 angstrom-width graphene channel at $\Delta T = 30$ K, reaching a flowrate at pressure difference up to $10^5$ bar in commercial RO processes and 230 bar in pressure-driven slip flow. All we have done is to provide a new idea of developing a high-efficiency desalination technology able to utilize various forms of energy.

# COMPUTATIONAL METHODS

**Molecular Dynamics Simulations.** Molecular model is shown in Figure 1. Two saline water reservoirs, at the opposite of ends of the simulation box, are divided by a graphene channel with the length of 10 nm. Period boundary conditions are utilized in the in-plane $x$ and $y$ directions. In $z$ direction, both sides of system are bounded by rigid pistons to apply pressure with same value and opposite direction. It aims to guarantee the water molecules totally entrance into the channel between graphene lamellas. Each saline water reservoir contains 4000 water molecules and 40 pair of $Na^+$ and $Cl^-$ (with the concentration of 0.56 M). The two graphene laminates in channel have the width of 3 nm in $y$ directions, and we consider five sets of interlayer spacing between two laminates, corresponding to 6.5 Å, 9 Å, 12 Å, 17 Å, and 22 Å. Simulations are firstly performed in the NVT ensemble lasting 500 ps for relaxation. After that, the system is switched to the NVE ensemble, and two specific regions at the opposite ends are thermostatic to perform a temperature gradient through the whole system. To avoid the spurious effect on water molecules entering into the channel, the thermostatic regions are set to begin at the position 4 nm away from the channel entrance, till the edge of the simulated box. The temperature differences between hot and cold reservoirs range from 30 K to 80 K at interval of 10 K, within the limits of 280 K to 350 K. Langevin thermostat is utilized to control the temperature of hot and cold water reservoir in our simulations. The time step to integrate the equation of motion is 0.5 fs. Simulations of thermal-driven flow are implemented after equilibration at average temperature of cold water and hot water using the Nosé-Hoover thermostat with a damping time constant of 50 fs. The number of water molecules and ions from two reservoirs are recorded every 5 ps to get converged results, and last for a sufficiently long time (20 ns) for data collection. The rigid simple point charge effective pair (SPC/E) model is used to describe the potential of water molecules. The interactions between water molecules and carbon atoms in graphene is considered as parameters $\sigma_{c-o}$ = 0.319 nm and $\varepsilon_{c-o}$ = 4.063 meV, which predicts a water contact angle for graphene corresponding to the value measured experimentally. The van der Waals interactions are truncated at 1.2 nm, and the long-range Coulomb interactions are computed by utilizing the particle-particle particle-mesh (PPPM) algorithm. All potential parameters are given in Table 1. The characteristic length $\sigma$ and energy parameter $\varepsilon$ between different atoms are employed by the common Lorentz-Berthelot combination rule. All MD simulations are employed using the large-scale atomic/molecular massively parallel simulator (LAMMPS) package[49]. The post-processing is made by Visual Molecular Dynamics (VMD)[50] and The Open Visualization Tool (OVITO)[51].

**Calculation of the Diffusive Coefficients.** To elucidate the diffuse behaviors of water confined inner graphene channels, we construct a sandwiched structure with water molecules intercalated between two graphene sheets. The size of graphene sheet is 10 nm × 3 nm, equal to the simulations of thermal-driven flow. Specified quantity of water molecules are placed between graphene laminations according to the molecular number inner thermal-driven channel, with $N_w$ from 329 to 1763. The variation of water molecules between graphene slabs with different $d$-spacing have been shown in Figure S12, and the averaged number of molecules in converged region is chosen as the specified value $N_w$. The molecular structures are equilibrated at NVT ensemble for 2 ns, and switched to NVE ensemble for data collection. The molecular diffusive coefficient $D$ can be obtained from the mean square displacement (MSD). It can be calculated from the long-time limit of MSD by $D_{msd} = \lim_{t \to \infty} \langle |\vec{r}(t) - \vec{r}(0)|^2 \rangle / 2d_i t$, where $d_i$ is the dimension of space, t is the simulation time, and $\langle \cdots \rangle$ is the ensemble average.

Table 1 Potential Parameters of Atoms

| atom | $\sigma$ (Å) | $\varepsilon$ (Kcal/mole) | charge($q$) |
|---|---|---|---|
| C(C-C) | 3.550 | 0.070 | 0 |
| O(H$_2$O) | 3.166 | 0.155 | -0.834 |
| H(H$_2$O) | 0.000 | 0.000 | 0.417 |
| Na$^+$ | 2.586 | 0.105 | 1 |
| Cl$^-$ | 4.402 | 0.105 | -1 |

**Acknowledgements**

The authors acknowledge the support of the National Natural Science Foundation of China (No. 51706157 and No. 51606082), the China Postdoctoral Science Foundation (No. 2017M612498 and No. 2018T110796), and the Natural Science Foundation of Hubei Province of China (No. 2018CFB470). The numerical calculations in this paper have been done on the supercomputing system in the Supercomputing Center of Wuhan University.

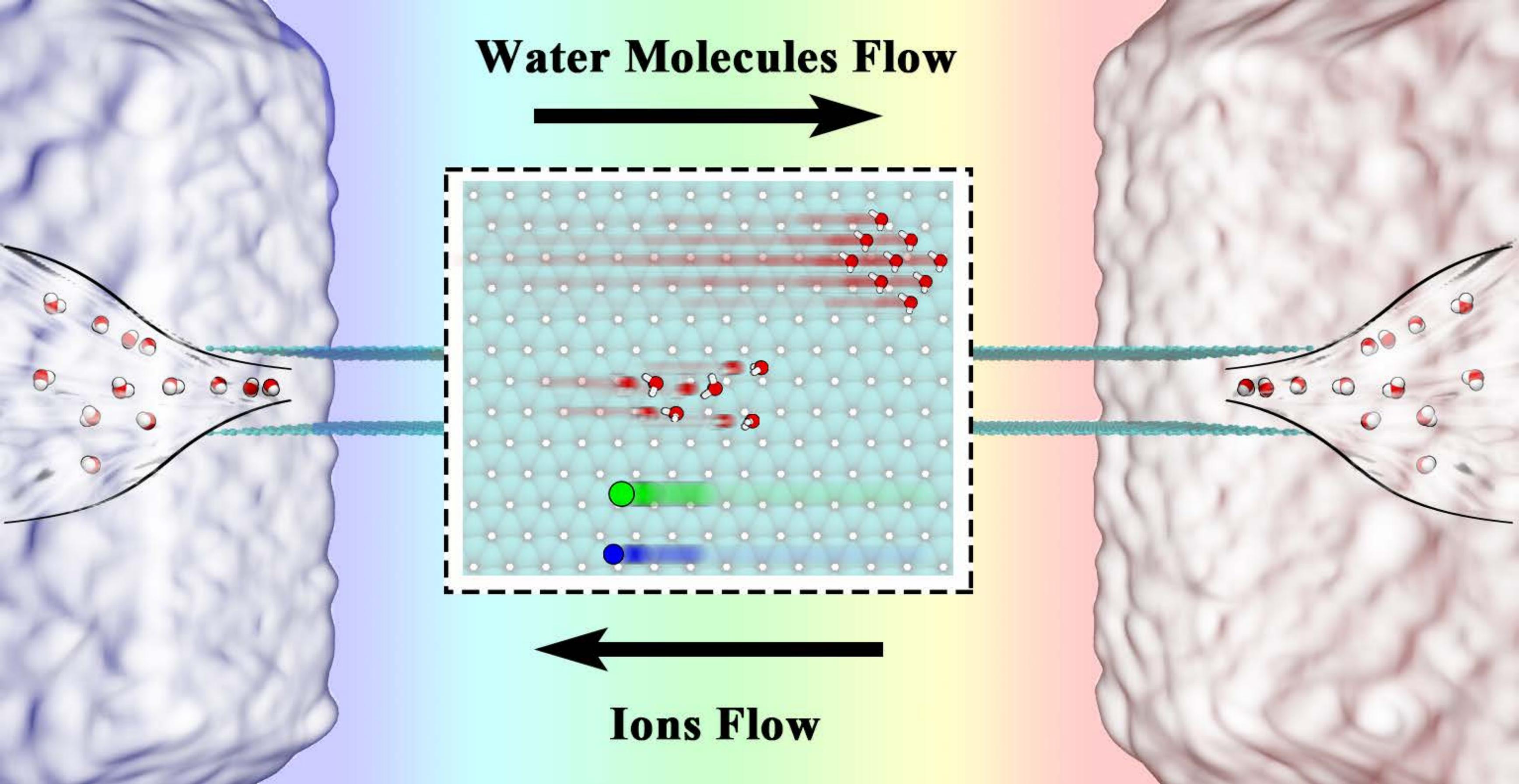

# Supplementary Information

# Thermal-driven Flow inside Graphene Channels for Water Desalination

Bo Chen, Haifeng Jiang*, Huidong Liu, Kang Liu, Xiang Liu, and Xuejiao Hu*

Key Laboratory of Hydraulic Machinery Transients (Wuhan University), Ministry of Education, School of Power and Mechanical Engineering, Wuhan University, Wuhan, Hubei 430072, China

*Corresponding authors. E-mail addresses: hfjiang@whu.edu.cn (H. Jiang), xjhu@whu.edu.cn (X. Hu).

## Contents

**1. Thermal Creep Flow**

**4. Calculations of the characteristics of nanoconfined water molecules**

Figure S13. The diagram of sandwiched structure with nanoconfined water molecules.

Figure S14. Pair correlation functions of atoms in water molecules.

Figure S15 Dependence of interaction energy between graphene slabs and per molecule to layer space.

**5. Temporal evolution of ions count at the system of graphene gallery with d-spacing of 6.5 Å**

Figure S16. Temporal evolution of ions count inner graphene channel and hot reservoir.

# 1. Thermal creep flow

The temperature field plays a significant role in flow for large Knudsen numbers, which is defined as the ratio of the molecular mean free path $\lambda$ and the characteristic length of the flow $L$, to characterize the degree of rarefaction. According to the rarified gas theory, a net flow can be induced along a temperature gradient on a diffusely reflecting wall. It is known as thermal creep flow. To analyze the problem in a simplified picture in Figure S1, we consider a patch of wall for investigation, which receives collisions from molecules originating from regions of different temperature. Since molecules arriving from the colder region carry a larger momentum than molecules originating from the colder side, a net momentum towards the colder side is imparted onto the wall from these molecules. Because the forces are mutual, a corresponding force in opposite direction acts on the molecules by the stationary wall. As a result, molecules move from the cold side to the cold side, with the corresponding flow being termed thermal creep. For momentum conservation, the velocity $u$ of thermal creep flow with a temperature gradient $\partial T/\partial x$ can be deduced by the kinetic theory of rarefied molecules, estimated as:

$$u = \frac{3+j}{4}\left(\frac{\eta}{\rho T}\frac{\partial T}{\partial x}\right)_s \quad (1)$$

Here, $j$, $\eta$, $\rho$, and $T_s$ is the internal degree of freedom, viscosity, density, and temperature of fluid molecules, respectively. For molecules flowing through a channel, we consider the temperature of entrance as $T_0$, so the temperature along the whole channel can be described as $T = T_0 + \frac{\partial T}{\partial x}x$. According to equation (1), the velocity $u$ is dependent on the molecular temperature near the surface, which is related to the position in channel along temperature gradient. Thus, the averaged velocity $u_m$ is introduced to assess the comprehensive effects on the thermal creep flow, defined as:

$$u_m = \frac{1}{L}\int_0^L \frac{3+j}{4}\left(\frac{\eta}{\rho(T_0+\frac{\partial T}{\partial x}x)}\frac{\partial T}{\partial x}\right)_s dx = \frac{3+j}{4}\cdot\frac{\eta}{\rho L}\cdot\ln(1+\frac{L}{T_0}\frac{\partial T}{\partial x}) \quad (2)$$

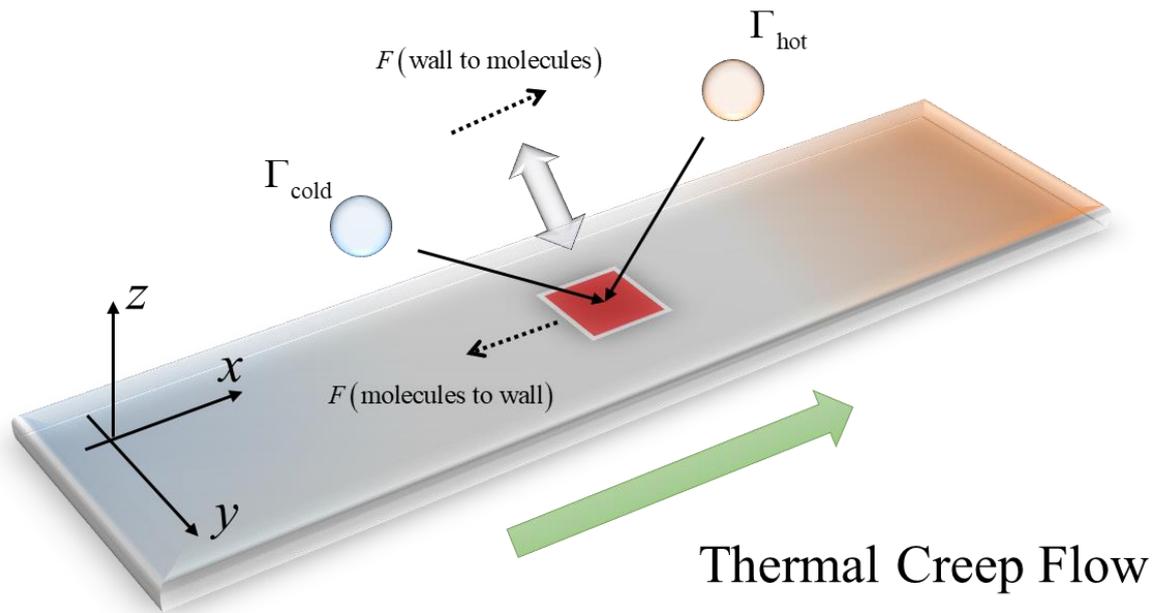

Figure S1. The simplified picture for explanation of thermal creep flow

## 2. Simulations of thermal-driven flow with variable temperature difference

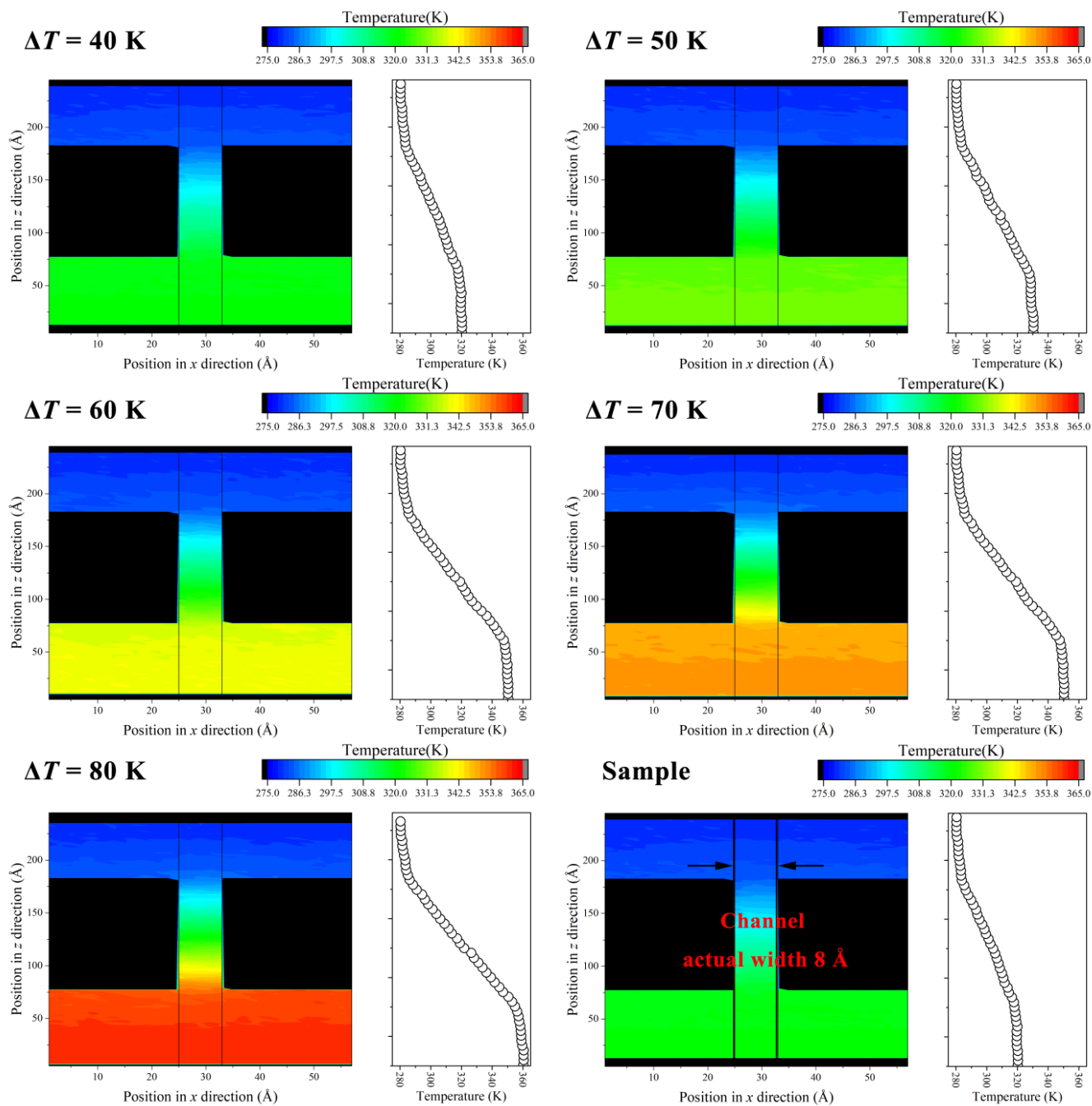

Figure S2. Temperature distributions of water flow in graphene channel with d-spacing of 12 Å on x-z plane (left panel) and along z direction (right panel). The temperature difference ranges from 40 K to 80 K.

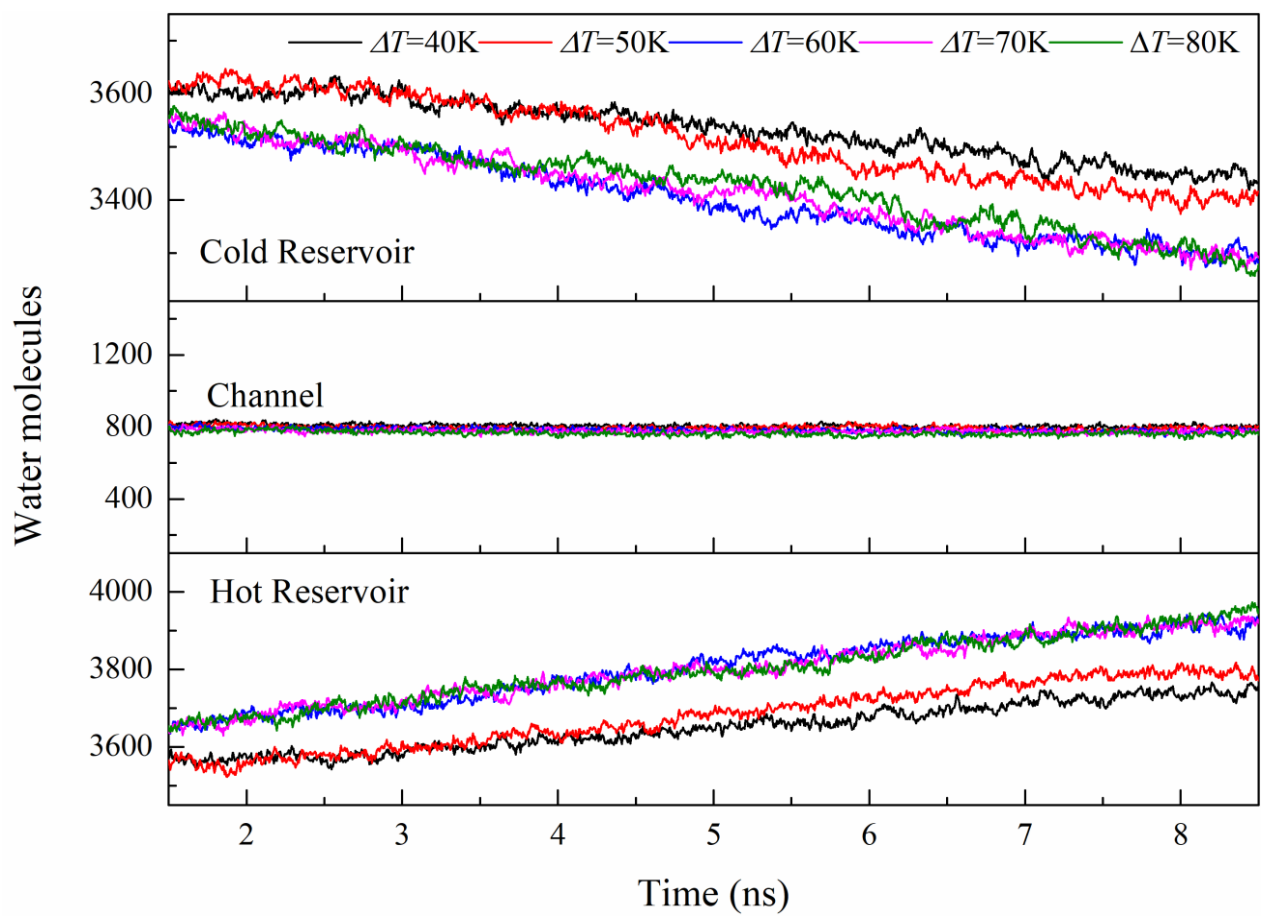

Figure S3. Temporal evolution of water molecules in cold reservoir (top panel), graphene channel (middle panel), and hot reservoir (bottom panel) with *d*-spacing of 12 Å. The temperature difference ranges from 40 K to 80 K.

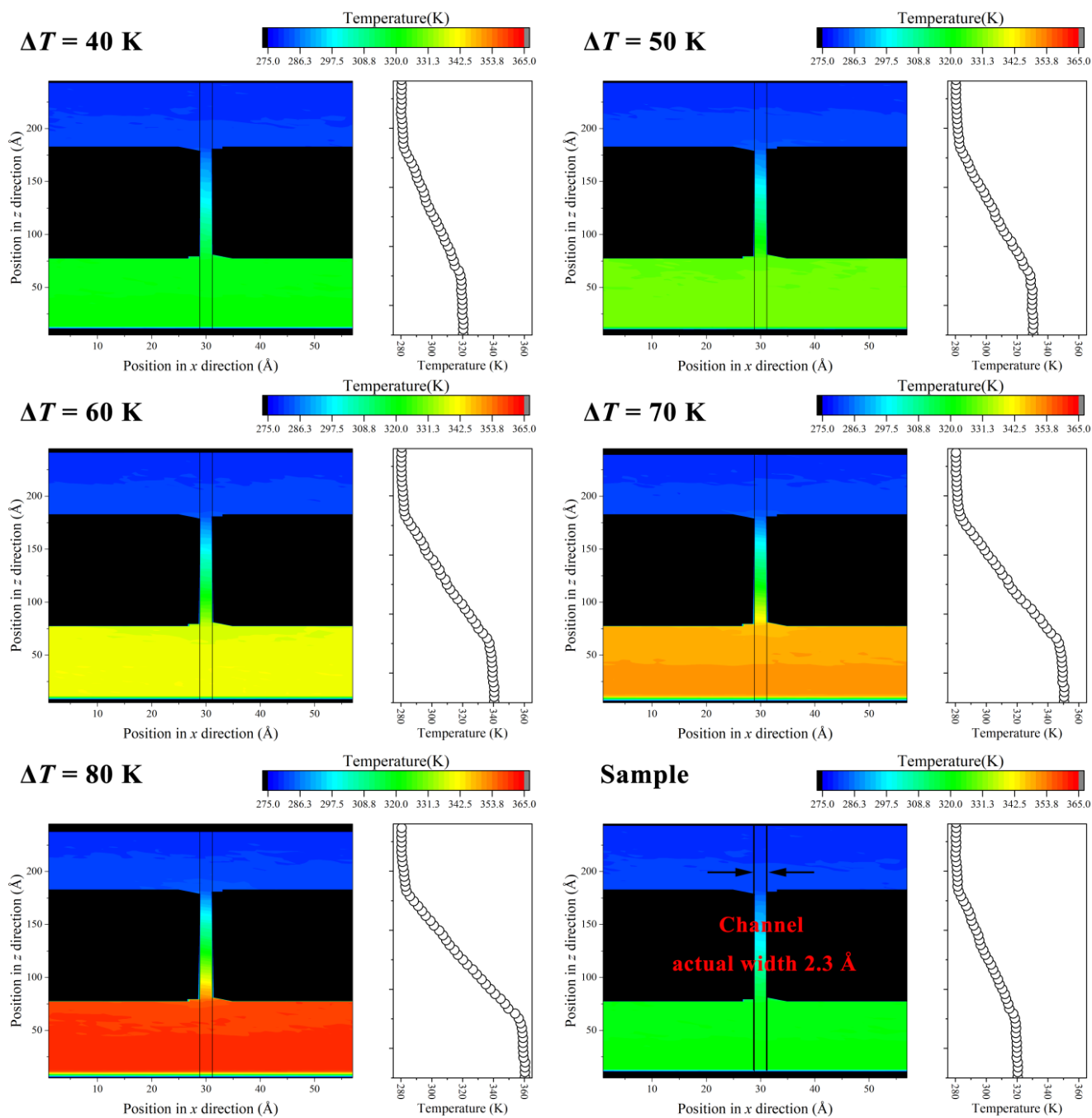

Figure S4. Temperature distributions of water flow in graphene channel with *d*-spacing of 6.5 Å on *x-z* plane (left panel) and along *z* direction (right panel). The temperature difference ranges from 40 K to 80 K.

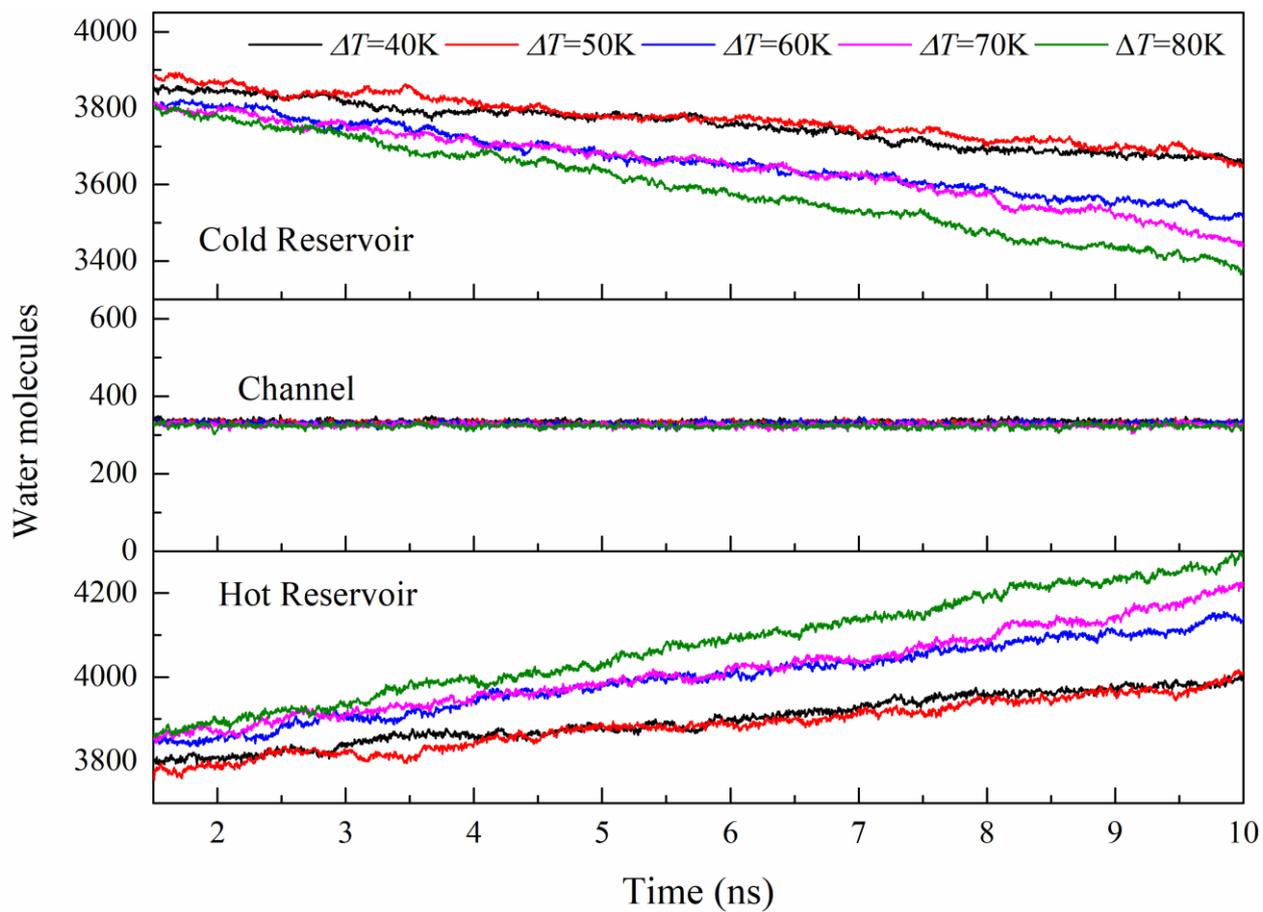

Figure S5. Temporal evolution of water molecules in cold reservoir (top panel), graphene channel (middle panel), and hot reservoir (bottom panel) with *d*-spacing of 6.5 Å. The temperature difference ranges from 40 K to 80 K.

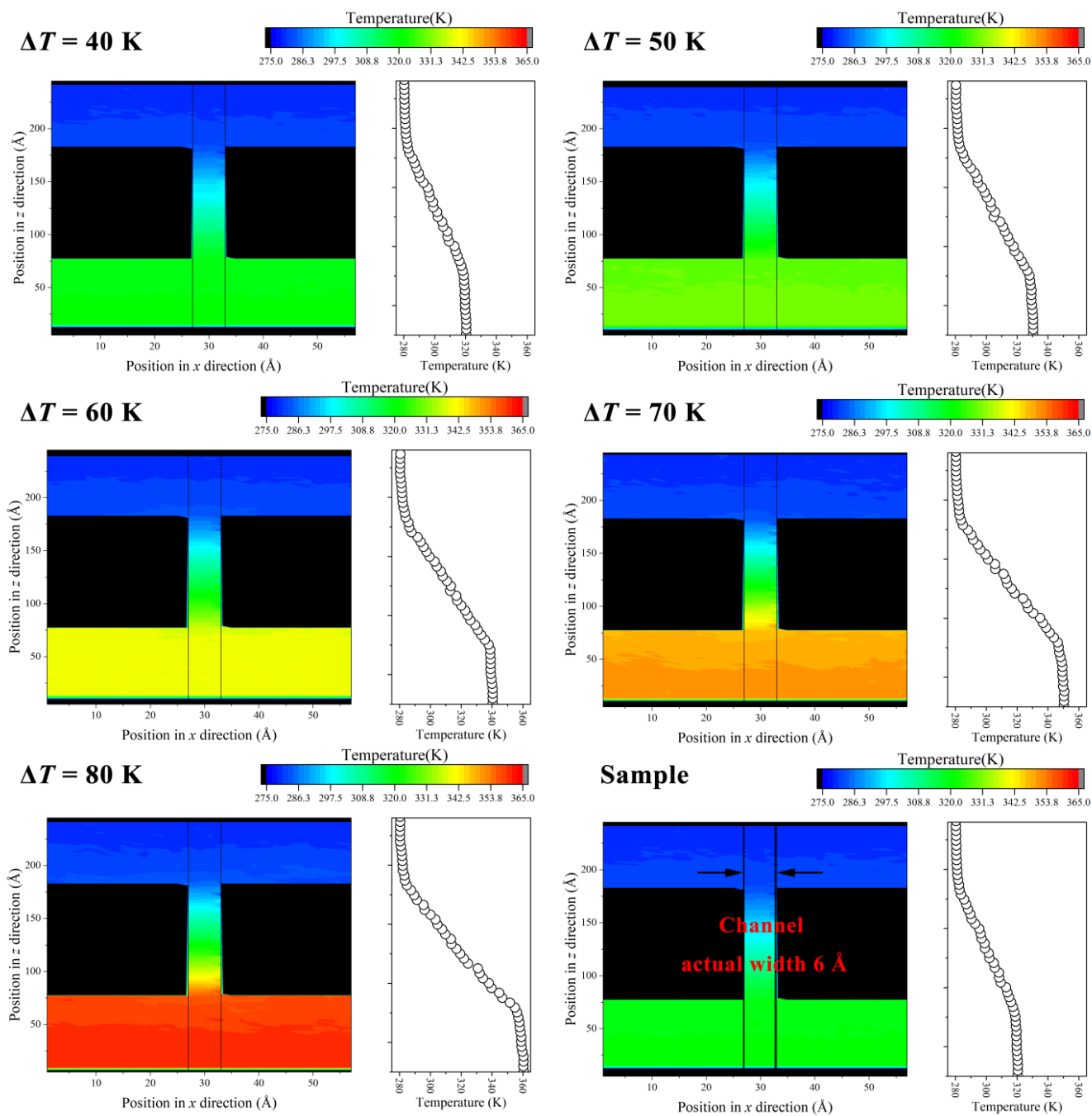

Figure S6. Temperature distributions of water flow in graphene channel with d-spacing of 9 Å on *x-z* plane (left panel) and along *z* direction (right panel). The temperature difference ranges from 40 K to 80 K.

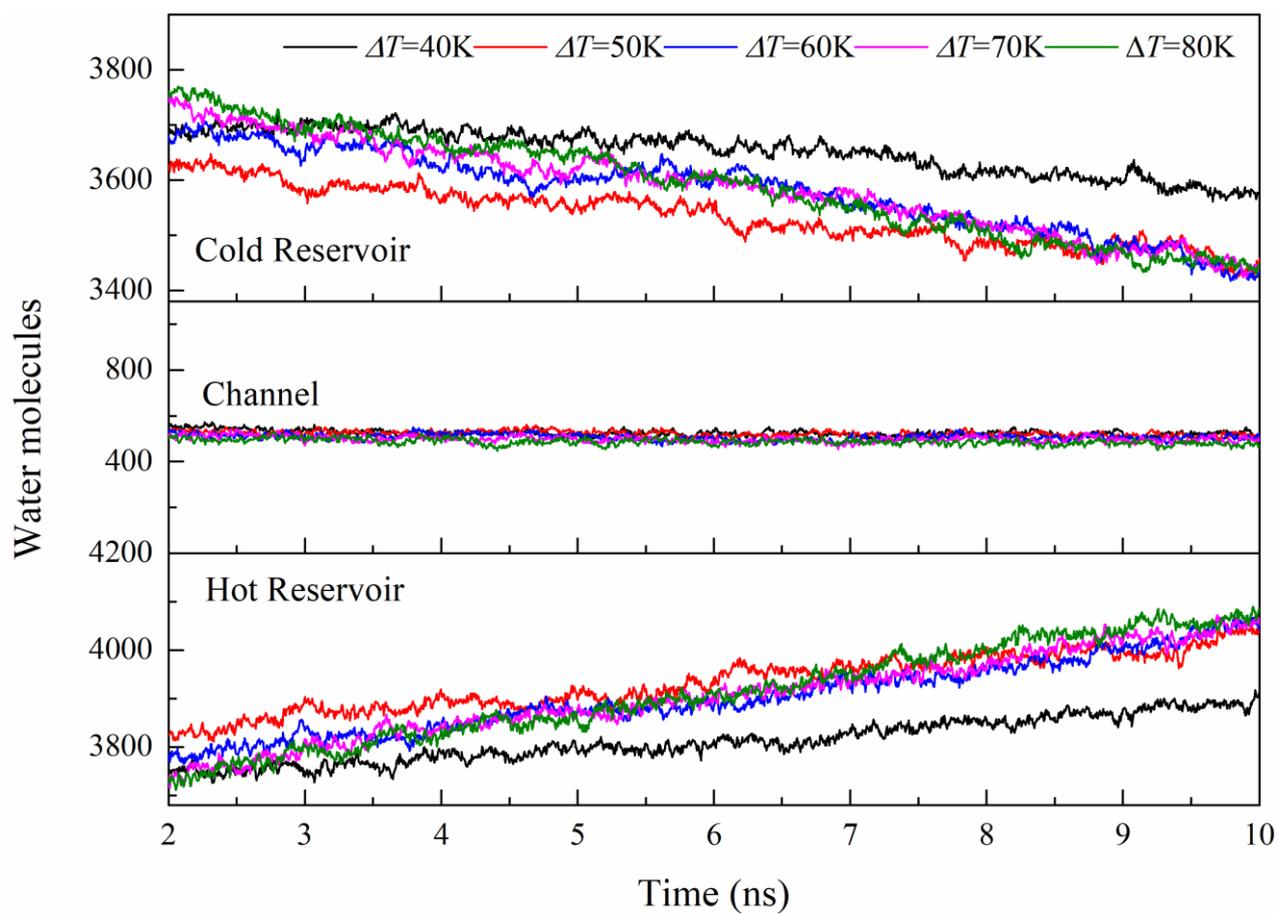

Figure S7. Temporal evolution of water molecules in cold reservoir (top panel), graphene channel (middle panel), and hot reservoir (bottom panel) with *d*-spacing of 9 Å. The temperature difference ranges from 40 K to 80 K.

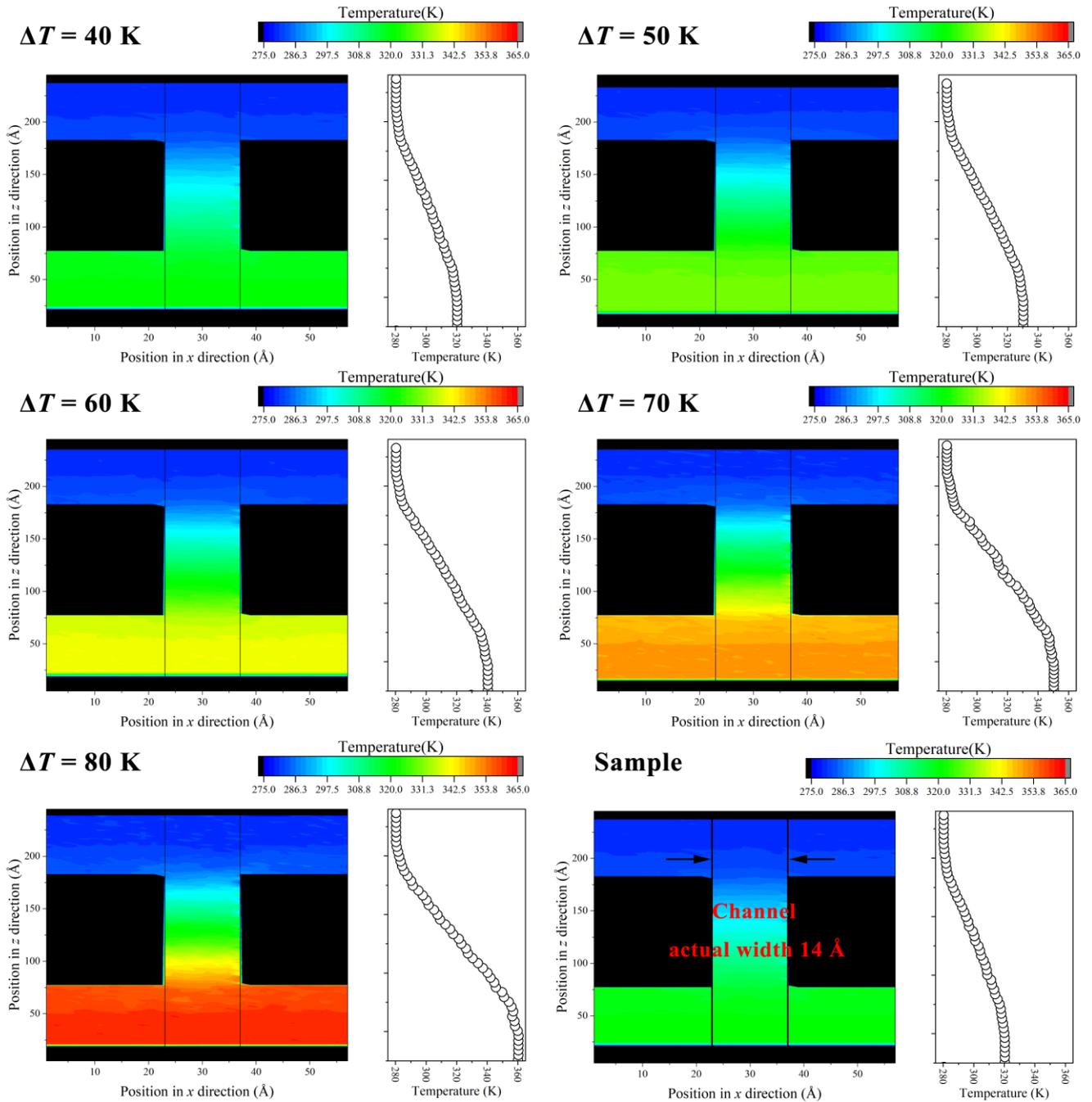

Figure S8. Temperature distributions of water flow in graphene channel with d-spacing of 17 Å on *x-z* plane (left panel) and along *z* direction (right panel). The temperature difference ranges from 40 K to 80 K.

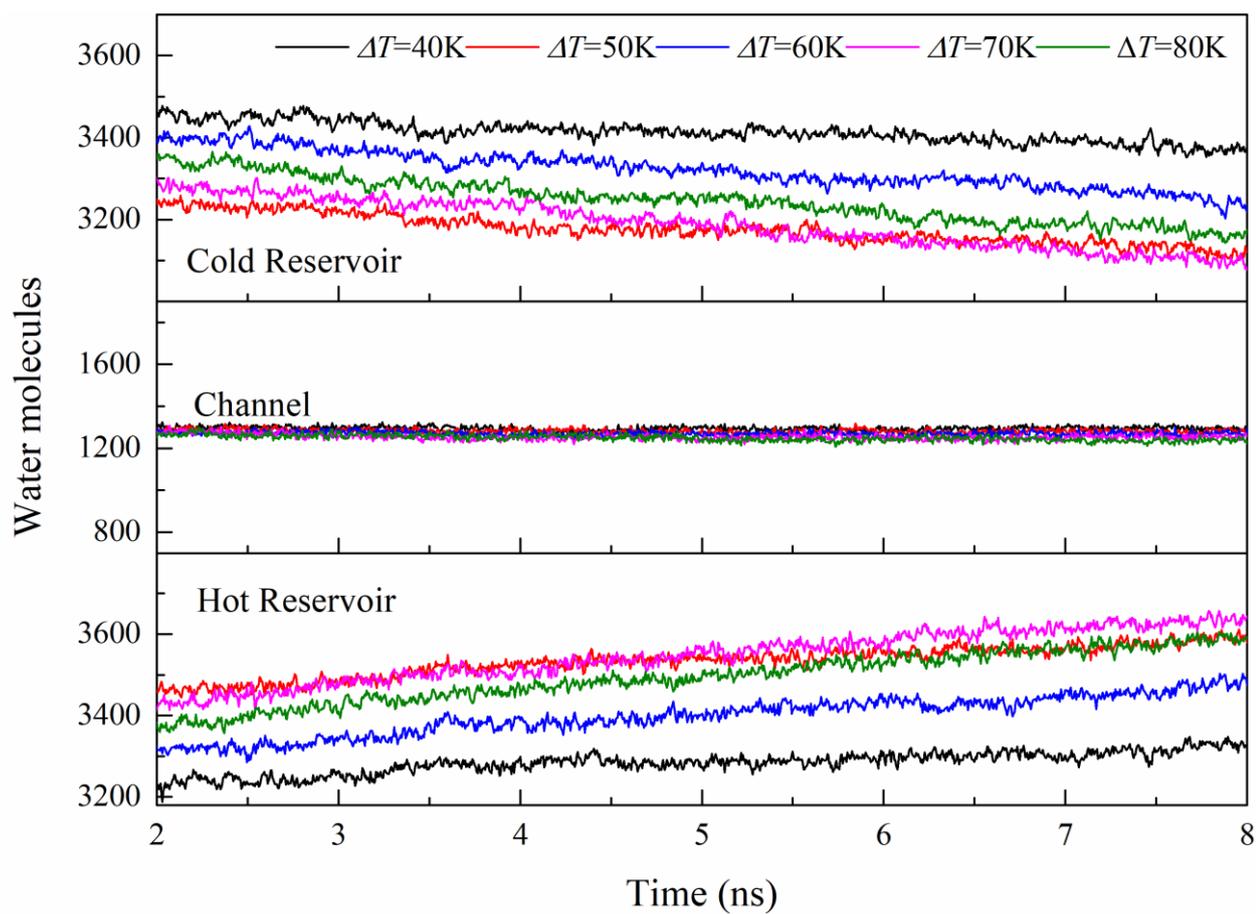

Figure S9. Temporal evolution of water molecules in cold reservoir (top panel), graphene channel (middle panel), and hot reservoir (bottom panel) with *d*-spacing of 17 Å. The temperature difference ranges from 40 K to 80 K.

## 3. Simulations of thermal-driven flow through graphene channel with different *d*-spacing

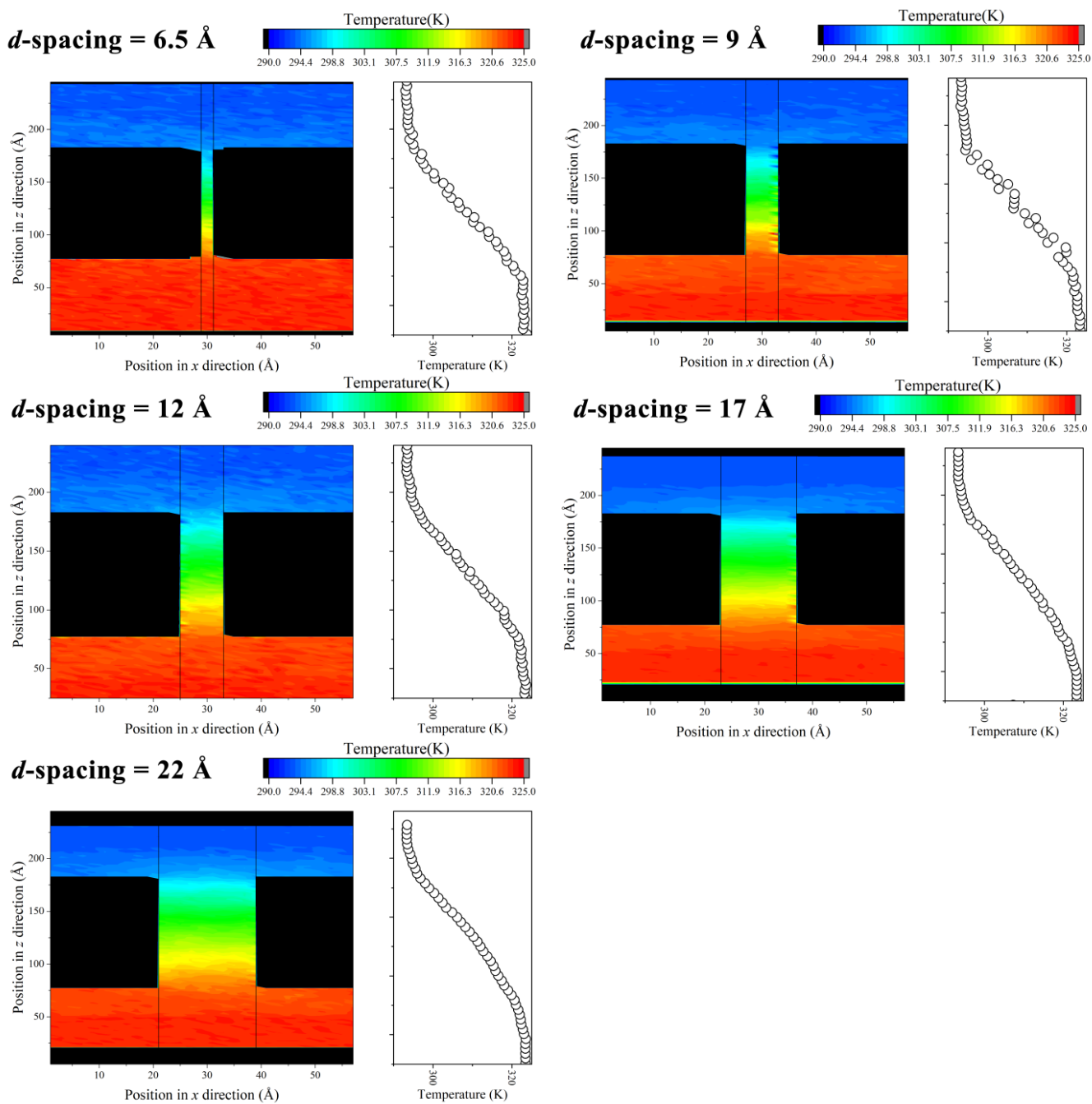

Figure S10. Temperature distributions of water flow in graphene channel with different *d*-spacing on *x-z* plane (left panel) and along *z* direction (right panel). The temperature difference is 30 K.

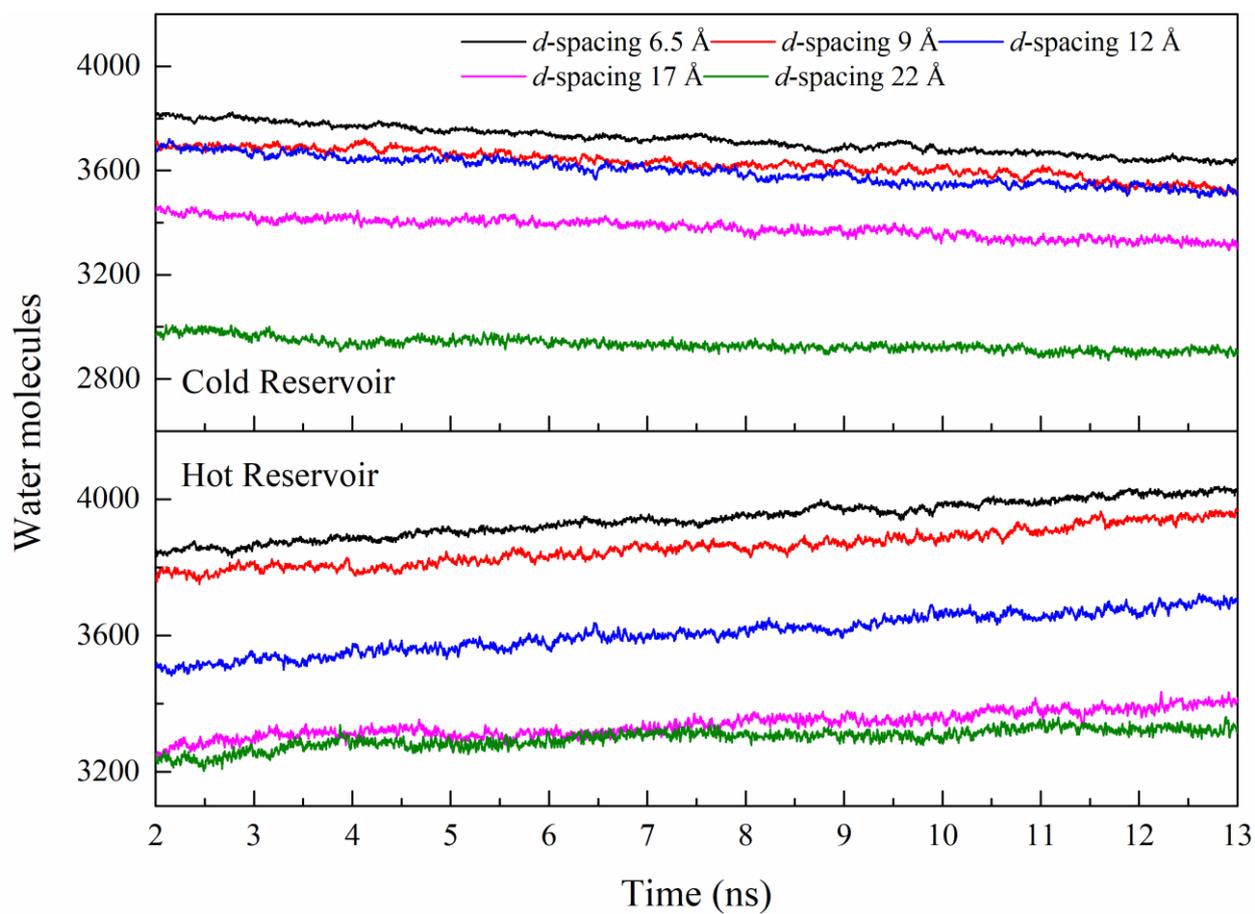

Figure S11. Temporal evolution of water molecules in cold reservoir (top panel), and hot reservoir (bottom panel) with different *d*-spacing. The temperature difference is 30 K.

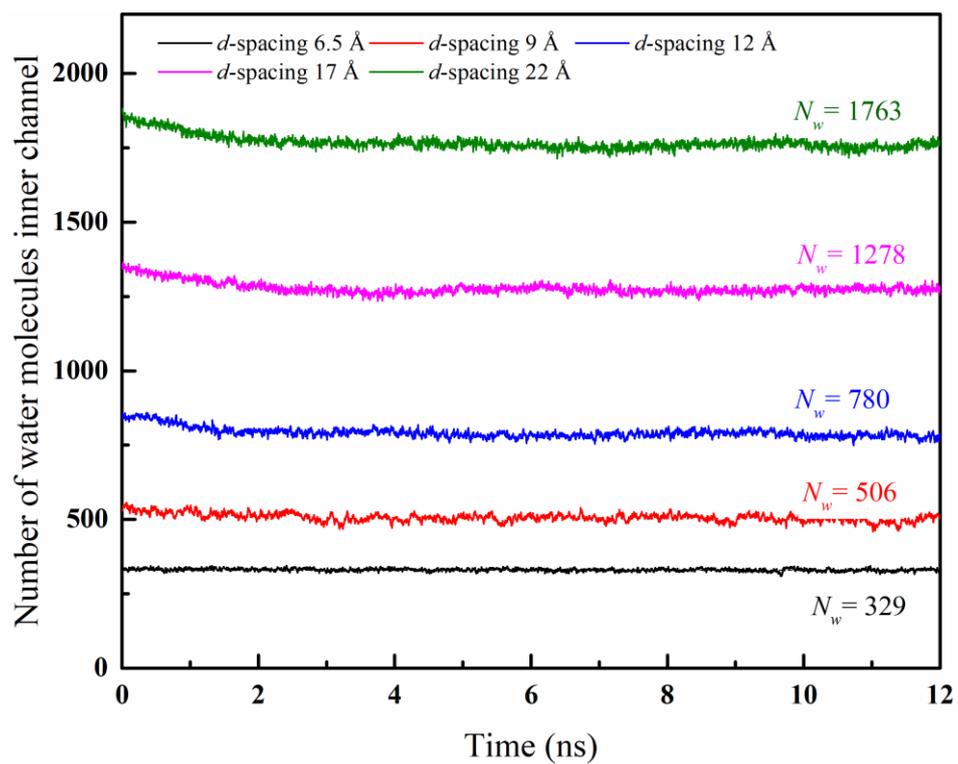

Figure S12. Temporal evolution of water molecules inner graphene channel with different *d*-spacing. The averaged number of water molecules $N_w$ have been calculated from 329 to 1763.

## 4. Calculations of the characteristics of nanoconfined water molecules

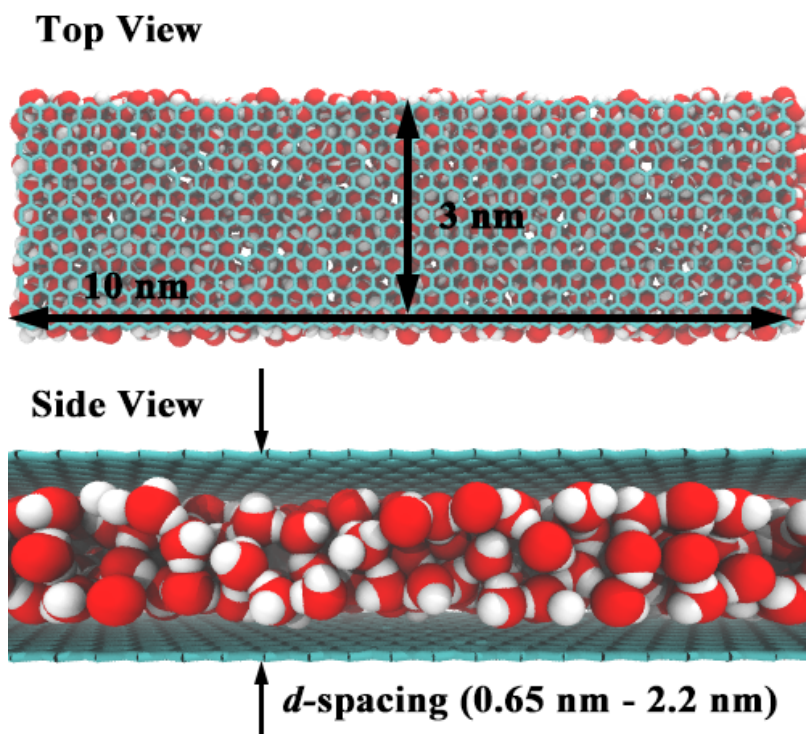

Figure S13. The diagram of sandwiched structure with water molecules intercalated between two graphene sheets. Red and white spheres represent oxygen atoms and hydrogen atoms. Cyan stick planes represent the graphene sheets.

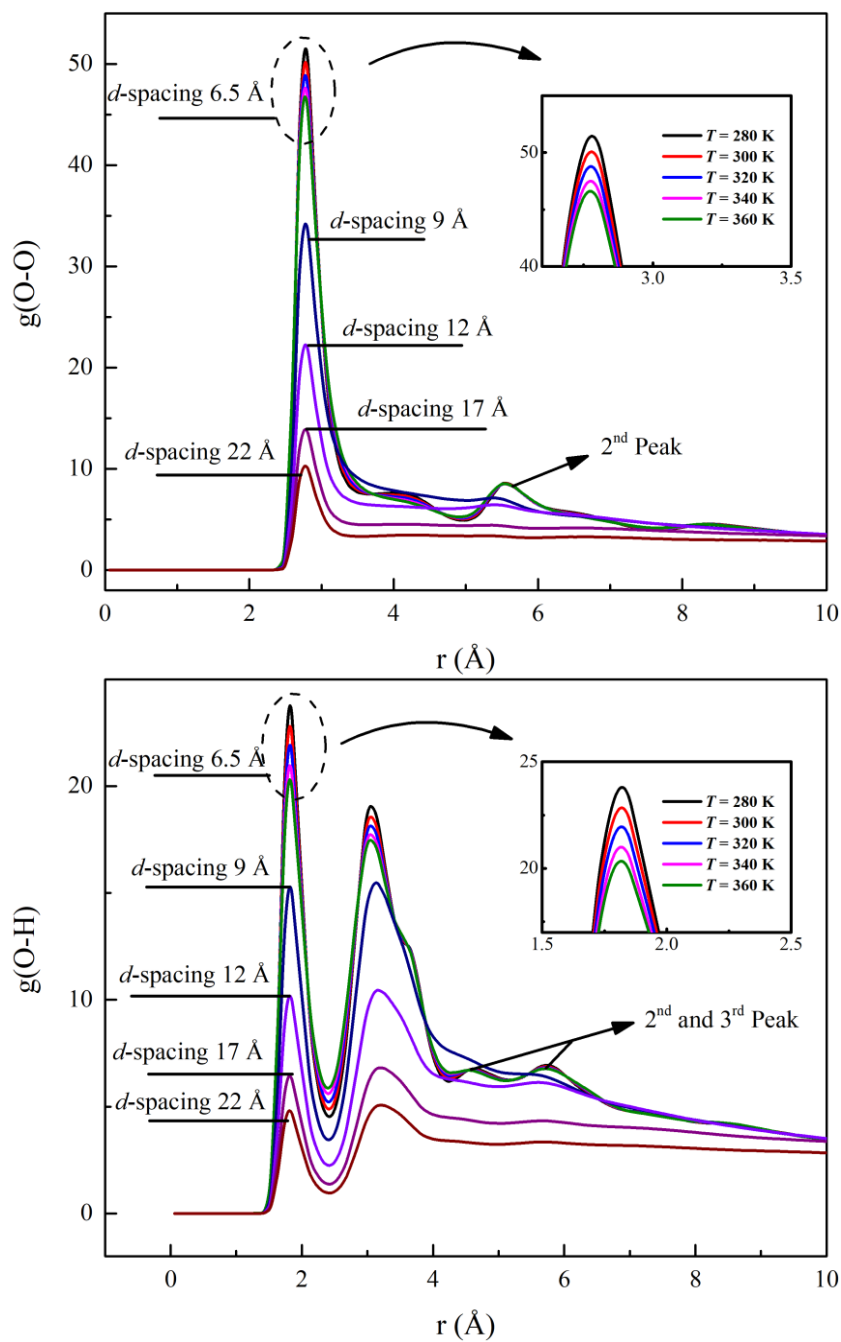

Figure S14 Pair correlation functions between O-O atoms (top panel) and O-H atoms (bottom panel) in water molecules intercalated between two graphene sheets. The *d*-spacing of graphene channels varies from 6.5 Å to 22 Å. The equilibrium temperature for channels with *d*-spacing beyond 6.5 Å is 300.15 K, and that at 6.5 Å ranges from 280 K to 360 K.

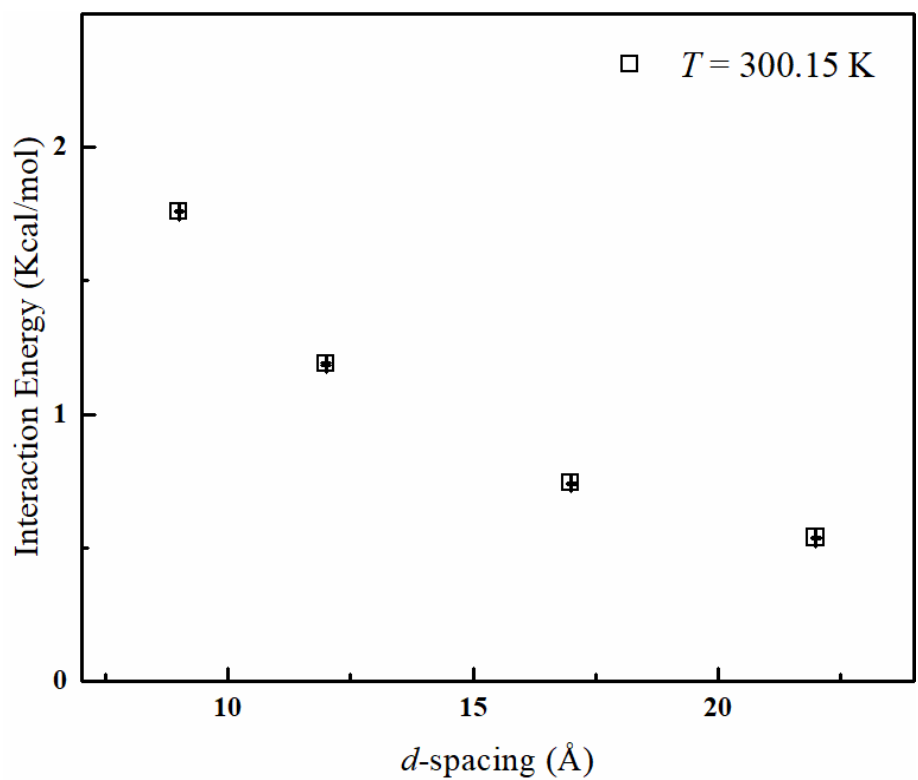

Figure S15. The dependence of interaction energy averaged to every intercalated water molecule to different layer spacing.

5. Temporal evolution of ions count at the system of graphene gallery with d-spacing of 6.5 Å

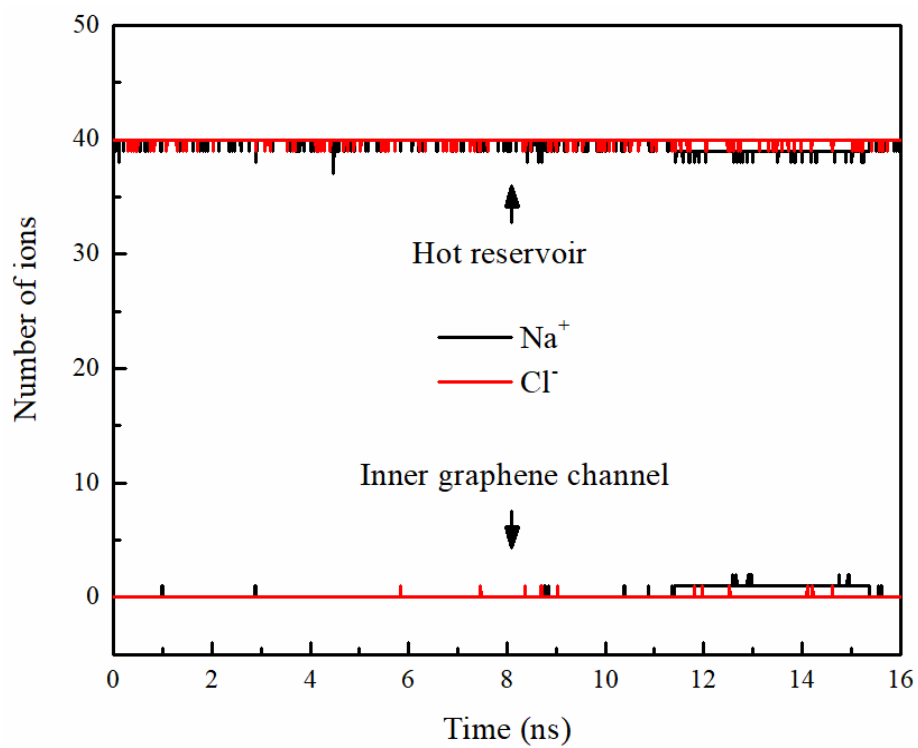

Figure S16. Temporal evolution of ions count inner graphene channel and hot reservoir at the thermal-driven system in graphene gallery with *d*-spacing of 6.5 Å.